**Resonances in non-relativistic free-free Gaunt factors with screened Coulomb interaction**

Ju Yan Wu,[1] Yong Wu,[2,3][*] Yue Ying Qi,[4] Jian Guo Wang,[2] R.K. Janev[5] and Song Bin Zhang,[1,†]

[1]School of Physics and Information Technology, Shaanxi Normal University, Xi'an 710119, China

[2]Institute of Applied Physics and Computational Mathematics, Beijing 100088, China

[3]Center for Applied Physics and Technology, Peking University, Beijing 100084, China

[4]School of Mathematics & Physics and Information Engineering, Jiaxing University, Jiaxing 314001, China

[5]Macedonian Academy of Sciences and Arts, PO Box 428, 1000, Skopje, Macedonia

**Abstract**

The effect of Coulomb interaction screening on non-relativistic free-free absorption is investigated by integrating the numerical continuum wave functions. The screened potential is taken to be in Debye-Hűckel (Yukawa) form with a screening length $D$. It is found that the values of the free-free Gaunt factors for different Debye screening lengths $D$ for a given initial electron energy $\varepsilon_i$ and absorbing photon energy $\omega$, generally lie between those of the pure Coulomb field and field-free case. However, for initial electron energies below 0.1 Ry and fixed photon energy, the Gaunt factors show dramatic enhancements (broad and narrow resonances) in the vicinities of the critical screening lengths, $D_{nl}$, at which the energies of $nl$ bound states in the potential merge into the continuum. These enhancements of the Gaunt factors can be significantly higher than their values in the unscreened (Coulomb) case over a broad range of $\varepsilon_i$. The observed broad and narrow resonances in the Gaunt factors are related to the temporary formation of weakly bound (virtual) and resonant (quasi-bound) states of the low-energy initial electron on the Debye-Hűckel potential when the screening length is in the vicinity of $D_{nl}$.

---

[*] wu_yong@iapcm.ac.cn

[†] song-bin.zhang@snnu.edu.cn

**I. Introduction**

The continuous photon emission and absorption processes, resulting from the free-free transitions of an electron in the field of a positive ion and caused by its acceleration in the field of the ion (known as bremsstrahlung and inverse bremsstrahlung), play an important role in a wide range of laboratory and astrophysical plasmas (plasma cooling, opacity, radiation transfer, etc.)[1,2]. The theoretical studies of these processes began in the early 1920s with their semi-classical description by Kramers [3] and Wentzel [4] and were followed by their full non-relativistic quantum-mechanical description by Gaunt [5] at the end of the same decade. Gaunt noticed that the classical result of Kramers differs from the quantum-mechanical one only by a factor, which is now known as the free-free Gaunt factor, $g_{ff}$.

For the case of an electron colliding with an isolated positive ion, the non-relativistic continuum wave functions of the continuum electron before and after collision are the Coulomb wave functions. With these wave functions analytic expressions for the non-relativistic Gaunt factors have been derived by Sommerfeld [6], Landau and Lifshitz [7] and Biedenharn [8], summarized in the acceleration gauge in [9]. Subsequently, numerical calculations of Gaunt factors have been performed and tabulated in various ranges of initial electron energy $\varepsilon_i$ and photon energy $\omega$ (see, e.g. [10-18]), including their thermal average. The most extensive $g_{ff}$ recent calculations are those in Ref. [11], covering the parameter space with $\log_{10}[\varepsilon_i(\text{Ry})] = -20$ to $+10$ and $\log_{10}[\omega(\text{Ry})] = -30$ to $+25$.

However, in many laboratory and astrophysical plasmas the many-body correlations of interacting charged particles introduce a collective screening effect on the Coulomb interaction and the motion of a continuum electron in the field of a positive ion cannot be described anymore by pure Coulomb wave functions. In the pair-wise approximation of the many-body correlation function, valid for weakly coupled classical plasmas, the screened Coulomb electron-ion interaction reduces to the Yukawa-type Debye-Hückel potential [19,20]

$$V(r) = -Ze^2 e^{-r/D} / r, \tag{1}$$

where $Z$ is the ionic charge, $D = (k_B T_e / 4\pi n_e)^{1/2}$ is the Debye screening length, $k_B$, $T_e$ and $n_e$ being the Boltzmann constant, and plasma electron temperature and density, respectively. Note that the weakly coupled (Debye) plasmas are defined by the condition $\Gamma = e/\bar{a}k_B T_e \ll 1$, where $\bar{a} = (3/4\pi n_e)^{1/3}$ is the inter-particle distance.

Comprehensive investigations of the atomic energy levels, atomic spectra, photon excitation and ionization, electron impact excitation and ionization, charge transfer and ionization by heavy particle collisions have been performed by many research groups in Debye plasmas revealing many new physics phenomena, as reviewed in [21]. The new physics in the radiative and collision processes involving the Debye-Hückel potential stems from its short-range character, lifting the $l$-degeneracy of Coulomb levels and supporting only a finite number of bound $nl$ states for any finite value of the screening length (see, e.g., [7]). The latter property of the potential implies that with decreasing $D$ the binding energies of $nl$ states decrease and at certain critical values $D_{nl}$ they successively enter into the continuum. In the small $D$-region around $D_{nl}$, the wave function of $nl$ state experiences a dramatic transformation which should be reflected in the transition probabilities of bound-free, free-bound and free-free processes for screening lengths close to $D_{nl}$.

In the present work, we shall study the free-free absorption in the field of the Debye-Hückel potential (1) in a broad range of initial electron and absorption photon energies, $\varepsilon_i$ and $\omega$. To the best of our knowledge, the only quantum-mechanical study (in the length gauge) for the free-free emission Gaunt factors in a Debye plasma is that of Lange and Schlüter [2] for the hydrogen ion ($Z$=1) and screening lengths $D$=10 a$_0$ and $D$=100 a$_0$ in the parametric range $\varepsilon_i$=0.2-4.0 Ry, $\omega$=0.0-2.0 Ry. For $D$=100 a$_0$, the values in the screened potential are almost indistinguishable from the Coulomb values, while for $D$=10 a$_0$ they differ from the Coulomb values, but show a smooth monotonic dependence on both $\varepsilon_i$ and $\omega$. In the present work, we have performed a comprehensive study of free-free absorption Gaunt factors in the ranges $\varepsilon_i$=10$^{-8}$-10$^2$ Ry and $\omega$=10$^{-7}$-10$^5$ Ry for different screening lengths. We have found that for a fixed photon energy $\omega$, the $\varepsilon_i$-dependence of free-free Gaunt factors can change dramatically when the screening length $D$ varies in the vicinity of the critical lengths $D_{nl}$ at which the bound $nl$ states in the Debye potential enter into the continuum, exhibiting broad resonances near $D_{ns}$ and narrow (shape-type) resonances near $D_{nl}$ ($l$>1) critical lengths. The physical origin of these resonant features will be discussed in detail in Section III. In the next section, we briefly present the computational methods and in section IV we give our conclusions. Atomic units (a.u.) will be used in the remaining part of this article, unless otherwise indicated explicitly.

## II. Computational methods

The theory of free-free absorption has been discussed in detail by Karzas and Latter [9]. With the initial and final state wave functions normalized in the energy space (number per unit volume per energy interval), the general expression of the non-relativistic free-free absorption cross-section for an ion with charge $Z$ in the acceleration gauge is given as [9]

$$d\sigma_e^{ab} = a_0^2 Z^2 \frac{64}{3} \left(\frac{e^2}{\hbar c}\right) \left[\left(\frac{Ry}{\hbar\omega}\right)^3 \frac{d(\hbar\omega)}{Ry}\right] k_i k_f \sum_{l=0}^{\infty} \left[(l+1)\tau_{l+1\leftarrow l}^2 + l\tau_{l-1\leftarrow l}^2\right], \quad (2)$$

where $\omega$ is the frequency of the absorbing photon, $\varepsilon_i$ and $\varepsilon_f = \varepsilon_i + \hbar\omega$ are the initial and final state electron energies with conjugate momenta $k_i$ and $k_f$, respectively. The free-free transition matrix element is $\tau_{l'\leftarrow l} = \int_0^\infty r^2 \psi_{l'}^f(k_f r) \frac{1}{r^2} \psi_l^i(k_i r) dr$, where $\psi_l^i$ and $\psi_l^f$ are the continuum wave functions of initial and final states, respectively, having the asymptotic behavior $\psi_l(kr) \stackrel{r\to\infty}{=} \sin(kr + \Delta_l)/(kr)$. The well-known classical Kramers result for the free-free absorption cross sections is [3]

$$d\sigma_K = a_0^2 Z^2 \frac{32\pi}{3\sqrt{3}} \left(\frac{e^2}{\hbar c}\right) \left[\left(\frac{Ry}{\hbar\omega}\right)^3 \frac{d(\hbar\omega)}{Ry}\right]. \quad (3)$$

The free-free Gaunt factor $g_{ff}$ is defined as the ratio between the quantal free-free absorption cross-section and the Kramers cross section [1,5]

$$g_{ff}(\varepsilon_i, \omega) = \frac{\sigma_e^{ab}}{\sigma_K} = \frac{2\sqrt{3}}{\pi} k_i k_f \sum_{l=0}^{\infty} \left[(l+1)\tau_{l+1\leftarrow l}^2 + l\tau_{l-1\leftarrow l}^2\right]. \quad (4)$$

For the pure Coulomb field, the continuum electron wave functions are the Coulomb wave functions and $g_{ff}$ can be analytically expressed in terms of complete hypergeometric functions $_2F_1(a,b,c,z)$ as [8]

$$g_{ff}(\varepsilon_i, \omega) = \frac{2\sqrt{3} I_0}{\pi} \left[\left(k_i^2 + k_f^2 + 2k_i^2 \eta_f^2\right)\frac{I_0}{k_i k_f} - 2k_i k_f \left(1 + \eta_i^2\right)^{1/2} \left(1 + \eta_f^2\right)^{1/2} \frac{I_1}{k_i k_f}\right], \quad (5)$$

where $\eta_i = Z/a_0 k_i$, $\eta_f = Z/a_0 k_f$, and $I_l$ ($l=0,1$) are defined as

$$I_l = \frac{1}{4}\left[\frac{4k_ik_f}{(k_i-k_f)^2}\right]^{l+1} \exp\left[\frac{\pi}{2}|\eta_i-\eta_f|\right] \frac{|\Gamma(l+1+i\eta_i)\Gamma(l+1+i\eta_f)|}{\Gamma(2l+2)} \times$$
$$\left|\frac{k_f-k_i}{k_f+k_i}\right|^{i\eta_i+i\eta_f} \times {}_2F_1\left[l+1-i\eta_f, l+1-i\eta_i, 2l+2; -\frac{4k_ik_f}{(k_i-k_f)^2}\right]. \quad (6)$$

We note that the Gaunt factor $g_{ff}$ remains finite also in the case $Z=0$ (or the field-free case), providing the limit when the effective charge of the screened ion tends to zero.

As mentioned earlier, in the Debye-Hückel potential (1) the continuum wave functions are obviously not anymore Coulomb waves and, since the radial Schrödinger equation with this potential cannot be solved analytically, the determination of the continuum wave functions has to be accomplished by direct numerical solution of radial Schrödinger equation with the potential (1). In the present work for this purpose we have employed the program RADIAL[22], which provides highly accurate numerical wave functions. Generally, the free-free transition matrix elements are calculated by partitioning the integration region into inner and asymptotic region; the numerical integrations are performed in the inner region and analytical expressions are used for the asymptotical region. However, if $\varepsilon_f$ is many orders of magnitude larger than $\varepsilon_i$, the number of mesh grids in the inner region for the final wave function can be many orders larger than that for the initial wave function. This results in a large inefficiency of the calculations. For such cases ($\varepsilon_f > 10^2 \varepsilon_i$), the inner region is partitioned into two sub-regions, so that the final wave function has reached into its asymptotic region in the outer sub-region, direct numerical integration is performed in the inner sub-region, while the numerical integration method for highly oscillating functions is employed in the outer sub-region.

Before presenting our computational results we mention that the radial Schrödinger equation with the potential (1) is scalable with respect to $Z$. Under the transformations: $\rho=Zr$, $\delta=ZD$, $\varepsilon(\delta)=\varepsilon(Z,D)/Z^2$, it reduces to the equation for the ion with $Z=1$. For the sake of simplicity, the notations for the energy and the screening length will be those of the unscaled case ($Z=1$). As mentioned in the Introduction, the most important changes in the Gaunt factors in the field of screened potential (1), when the screening length varies, take place in the vicinity of critical screening lengths $D_{nl}$. In Table 1 we display the screening lengths for the $nl$ states with $n \leq 6$, taken from Ref. [23].

## III. Results and discussion

For the purpose of discussing the general properties of non-relativistic Gaunt factors, we display in Fig. 1 the Gaunt factors for the pure Coulomb case ($Z=1$) and the field-free case ($Z=0$) in the ranges $\varepsilon_i=10^{-8}$-$10^8$ Ry, $\omega=10^{-7}$-$10^7$ Ry. The figure shows that at high $\varepsilon_i$ and a fixed value of $\omega$ both Gaunt factors increase with increasing $\varepsilon_i$ and for sufficiently large values of $\varepsilon_i$ the Coulomb Gaunt factors gradually approach those of the field-free case as the electron energy becomes much larger than the potential energy of the ion. The increase of $g_{ff}$ with increasing $\varepsilon_i$ is due to the increased number of $l$-waves contributing to the sum in Eq. (4). For a given $\omega$, and when $\varepsilon_i$ is very small, the Gaunt factor is completely dominated by the $s$-wave contribution. Consequently, in the Coulomb case $g_{ff}$ is constant, while in the field-free case it is proportional to $k_i = \sqrt{2\varepsilon_i}$, in accordance with the Wigner threshold law [24]. Fig. 1 also shows that for a given $\varepsilon_i$, $g_{ff}$ decreases with the increase of $\omega$, resulting from the fact that for a continuum electron it is more difficult to absorb a high energy photon. This follows also from the classical Kramers free-free absorption cross section, $\sigma_K \propto \omega^{-3}$ (cf. Eq. (3)).

Bearing in mind that the field-free and pure Coulomb cases are the limiting cases of screened potential (1) ($D \to 0$ and $D \to \infty$, respectively), Fig.1 indicates the regions in the ($\varepsilon_i$, $\omega$) parametric plane in which the Gaunt factors in the screened case differ from the Coulomb ones. From physics point of view, when analyzing the behavior of Gaunt factors $g_{ff}$ for the potential (1), the most interesting are the regions $\omega \gg \varepsilon_i$ and $\omega \sim \varepsilon_i$, with $\varepsilon_i <1.0$ Ry. In the first case the wave function of incident electron is sensitive to the bound state structure of the potential (due to the longer time the electron spends in the field), while after absorbing an energetic photon it becomes insensitive to the potential (the high-energy electron quickly leaves the attractive field). In the second case the electron both before and after absorbing a low-energy photon has wave functions that are sensitive to the bound state structure of the potential. In what fallows we shall analyze both these two physical situations.

In Fig. 2 we show the variation of $g_{ff}(\varepsilon_i, \omega)$ in the ranges $\varepsilon_i=10^{-8}$-$10^2$ Ry and $\omega=10^{-7}$-$10^5$ Ry for the pairs of screening lengths $D=0.8, 0.9$ a.u., $D=3.2, 3.3$ a.u. and $D=4.4, 4.6$ a.u. in the vicinities of critical screening lengths $D_{1s}=0.840$ a.u., $D_{2s}=3.223$ a.u. and $D_{2p}=4.541$ a.u., respectively (panels (a), (b) and (c), respectively). We note that the first value of these $D$-pairs is smaller than the corresponding critical screening length $D_{nl}$ (i.e., the $nl$ state is already in the continuum), while the second value of the pair is larger than the corresponding $D_{nl}$ (the $nl$ state is still bound in the potential). The figure shows that in the regions around $\varepsilon_i \sim 10^{-3}$, $\sim 10^{-5}$, $\sim 10^{-3}$ Ry in the panels (a), (b), (c), respectively, the Gaunt factors for both $D$-values around the critical screening lengths $D_{1s}$ and $D_{2s}$ exhibit dramatic enhancements, but in the vicinity of $D_{2p}$ only for $D$=4.5 a.u. such enhancement is observed. It should also be noted that Gaunt factor enhancements related to the critical screening length $D_{1s}$ appear at about a factor of ten higher energies than those related to $D_{2s}$. This is a consequence of the stronger screening of the ion for smaller $D$.

Fig. 2 also shows the variation of magnitudes of Gaunt factors with $\omega$ in the considered ($\varepsilon_i$, $\omega$) domain, including those of the enhancement peaks. Outside the enhancement regions $\Delta\varepsilon_i$, the Gaunt factors decrease with increasing $\omega$, like in the field-free and Coulomb cases. The Gaunt factor enhancement peaks for the selected values of $\omega$ maximize for $\omega$=10 Ry, and then decrease for larger or smaller $\omega$ value. It should be remarked that for $\omega$=$10^{-5}$ Ry and $\omega$=$10^{-7}$ Ry, the peaks for the screening length $D$=4.5 a.u. are converted into dips.

The broad Gaunt factor enhancements for the screening lengths near the critical $D_{ns}$ screening lengths can be understood on the basis of the general theory of low-energy particle scattering on a short-range potential [25]. The low-energy electron, when scattering on a short-range potential (1), can form either near-zero-energy bound (for $D$ slightly above $D_{ns}$) or a virtual (for $D$ slightly below $D_{ns}$, i.e. with positive energy) $s$-states, the wave functions of which have both large amplitudes and radial extensions [25]. In the scattering problem, the formation of such states leads to a large cross section enhancement (called in nuclear physics zero-energy or broad resonance [26]). The height and half-width of a broad resonance for a given $D$ depend only on the distance $|D_{ns} - D|$ and not on the character of the near-zero energy state. As panels (a) and (b) of Fig. 2 show, the peaks of Gaunt factors for $D$=0.8 a.u. and $D$=3.2 a.u., that are respectively closer to the screening lengths $D_{1s}$ and $D_{2s}$ have higher peaks and larger half-widths than their co-partners $D$=0.9 a.u. and $D$=3.3 a.u.. Their energy positions are also lower than those for $D$=0.9 a.u. and $D$=3.3 a.u., respectively.

The narrow resonant features in the Gaunt factors in the panel (c) of Fig.2 for the screening length $D$=4.5 a.u., that is slightly below the critical screening length $D_{2p}$=4.541 a.u., have obviously a different character than the broad resonances in the panels (a) and (b). The origin of these peak structures is the temporary capture of the electron's $p$-wave in the effective potential (containing an $l=1$ centrifugal barrier) when the electron energy coincides with the energy of a quasi-bound state formed when the bound $2p$ state enters the continuum at $D=D_{2p}$ (shape resonance). As seen from Fig. 2c, for the screening length $D=4.6$ a.u. such shape resonance does not appear in the Gaunt factor as for this value of $D$ the $2p$ state is still bound (although rather weakly). As mentioned earlier for $\omega=10^{-5}$ Ry and $\omega=10^{-7}$ Ry, these resonances manifest themselves as dips in the Gaunt factors (see, e.g., [25]). Note that the quasi-bound states in the Debye-Hückel potential (1) have been studied in [27] for $l$ up to $l=10$, providing also an estimate of the range of $D$ below $D_{nl}$ in which quasi-bound states can be formed, as well as the values of their half-widths.

A further insight in the physics of free-free absorption in a Debye plasma can be gained from Fig.3, where the contributions of $s$- and $p$- partial waves to the Gaunt factors for values of $D$ in the vicinity of $D_{2s}$ and $D_{2p}$ are shown for the typical $\omega=10$ Ry (panels (b) and (d)), together with the phase shifts of $s$- and $p$- partial waves (panels (a) and (c)). In the panels (b) and (d) we also show the contribution to the Gaunt factors from all $l>1$ partial waves. The dominance of $s$-wave contribution to the Gaunt factors for $D=3.2$ a.u. and $D=3.3$ a.u. in panel (b) in the energy region below 10 Ry is evident. The Gaunt factor involving the virtual $s$-state formed for $D=3.2$ a.u. is larger than the one involving the near-zero-energy bound $s$-state formed for $D=3.3$ a.u.. From the panels (a)-(d) we see that both phase shifts and Gaunt factors of $p$-partial waves are insensitive to the variation of $D$ in the vicinity of $D_{ns}$ critical screening lengths.

In the panels (c) and (d) of Fig. 3, the phase shifts and Gaunt factors for the $s$- and $p$- partial waves are shown, respectively, for $D=4.5$ a.u. and $D=4.6$ a.u.. Remarkable features in these panels are the jump of the $p$-wave phase shift by $\pi$ radians at $\varepsilon_i \sim 10^{-3}$ Ry and the related resonance at the same energy in the $p$-wave contribution to the Gaunt factor for $D=4.5$ a.u.. For $D=4.6$ a.u., which is larger than $D_{2p}=4.451$ a.u., both the $p$-wave phase shift and Gaunt factor show a smooth variation in the entire energy range. It should be remarked that the $s$-wave Gaunt factors for $D=4.5$ a.u. and $D=4.6$ a.u. are indistinguishable in the considered energy range and in the region below ~80 Ry they are larger than the corresponding ones for the $p$-wave, except for the shape resonance for $D=4.5$ a.u. in the relatively narrow region around $\varepsilon_i \sim 10^{-3}$ Ry. It is also worthwhile to note that all *l*-wave Gaunt factors shown in the panels of Fig. 3 obey the Wigner $\varepsilon_i^{l+1/2}$ threshold law.

A more complete picture of the energy behavior of Gaunt factors in the energy range $\varepsilon_i=10^{-8}$-$10^2$ Ry for the typical $\omega=10$ Ry for screening lengths in the ranges $D$=6.0-7.5 a.u., $D$=8.0-8.8 a.u., $D$=10.0-11.0 a.u. around the critical screening lengths $D_{3s}$=7.172 a.u., $D_{3p}$=8.872 a.u. and $D_{3d}$=10.947 a.u. in the panels (a), (b) and (c) of Fig.4, respectively. Also shown are the Gaunt factors in the pure Coulomb field and field-free cases. The figure illustrates the energy ranges within which, due to the resonance enhancements, the Gaunt factors in the screened potential (1) can become significantly larger than in the pure Coulomb case.

The broad *s*-resonances in panel (a) appear in the region below ~0.1 Ry. The magnitude of the *s*-resonance is larger, the closer its screening length is to $D_{3s}$. This rule is valid also for the shape resonances. For $D$=6 a.u. the potential cannot support a virtual *s*-state and a broad resonance cannot be formed. The shape resonances for $D$ around $D_{3p}$ are located in the energy range $5\times10^{-5} - 5\times10^{-3}$ Ry (cf. panel (b)). Resonances are formed only for $D$=8.8, 8.7 and 8.5 a.u. with peak values decreasing with increasing the difference $D_{3p}-D$. For $D$=8.9 a.u. and for $D$=8.0 a.u. shape resonances do not appear in the Gaunt factor energy dependence, since for $D$=8.9 a.u. the 2*p* state is still bound and for $D$=8.0 a.u. the electron energy is above the top of centrifugal barrier of the effective potential. The shape resonances for $D$=10.9, 10.8 and 10.6 a.u., associated with the crossover of 3*d* state into the continuum, are located in the energy range $5\times10^{-4} - 5\times10^{-2}$ Ry (cf. panel (c)) but only the resonance for $D$=10.9 a.u. in a small energy range has values above those in the Coulomb case. The half-widths of *d*-wave shape resonances are significantly smaller than those for the *p*-wave shape resonances in panel (b).

The Gaunt factors exhibit the same or similar patterns for the screening lengths $D$ around critical $D_{ns}$ and $D_{nl}$ (*l*>1) critical screening lengths at least with *n* up to 6. In Fig. 5 we show the Gaunt factors for pairs of selected screening lengths around the critical screening lengths $D_{6s}$=28.427 a.u., $D_{6p}$=31.080 a.u., $D_{6d}$=34.286 a.u., $D_{6f}$=37.950 a.u., $D_{6g}$=42.018 a.u. and $D_{6h}$=46.459 a.u. (panels (a) to (f) respectively) to cover a broad range of *D*. The values of *D*-pairs around each critical screening length $D_{nl}$ are chosen such that one of them is smaller and the other is larger than $D_{nl}$. The Gaunt factors for the *D*-pair near $D_{6s}$ both exhibit broad resonances (panel (a)) since for $D$=28.4 a.u. and $D$=28.5 a.u. the potential supports a virual and a weakly bound state, respectively. For all other $D$-pairs only the Gaunt factor for the smaller $D$ value exhibits a shape resonance as for the other $D$ value the $nl$ state in the potential is bound. The two broad resonances in the panel (d) appear due to the fact that the critical screening length for the 7s bound state in the potential has the value $D_{7s}$=38.64 a.u. [28] (close to $D_{6f}$=37.950 a.u.) and for the screening lengths $D$=37.9 a.u. and $D$=38.0 a.u. the potential supports virtual states. Thus, the Gaunt factor for $D$=37.9 a.u. in the panel (d) shows two overlapping (broad and shape) resonances, while the broad resonance for $D$=38.0 a.u. is due to the s-wave virtual state. Due to the large half-width of s-wave (broad) resonances their overlap with the shape ($l$>0) resonances should not be a so rare event in the region of large $D$. Whenever a pair of $D_{ns}$ and $D_{n'l'}$ ($l'$>0) are close enough to each other and $D$ is close to both of them (but on the left side of $D_{n'l'}$) an overlap of the s- and $l'$-resonances can appear. From Table 1 we can see that the pairs $D_{5s}$–$D_{4d}$ and $D_{6s}$–$D_{5f}$ provide conditions for such resonance overlapping. Due to their small widths the overlap of two shape resonances should be considered as a very rare event.

In closing this section we note that similar resonant structures with the same physics origin have been observed in the cross sections for photoionization of hydrogen-like ions [23,29] and negative hydrogen ion [30], low-energy electron impact excitation of hydrogen atom near the $n$=2 and $n$=3 thresholds [31-33] and fast-electron impact ionization of hydrogen-like ions [34,35] processes in Debye plasmas.

**IV. Conclusions**

In the present work we have investigated the properties of non-relativistic free-free absorption Gaunt factors $g_{ff}(\varepsilon_i, \omega)$ in the field of the screened Yukawa (Debye-Hűckel) potential in wide ranges of incident electron energy $\varepsilon_i$, absorbed photon energy $\omega$, and the screening length $D$ of the potential. We have revealed that the energy dependence of Gaunt factors for a given photon absorption energy $\omega$ is not a smooth increasing function like in the pure Coulomb interaction case, but exhibits resonant structures for screening lengths in the vicinity of critical screening lengths $D_{nl}$ at which the bound $nl$ states in the screened potential merge into the continuum. For screening lengths $D$ in the vicinity of $D_{ns}$, these structures are broad resonances related to the near-zero-energy virtual (for $D$<$D_{ns}$) and bound ($D$>$D_{ns}$) states. For $D$ in the left vicinity of $D_{nl}$ ($l$>0), the enhancement structures have shape-type resonance character, related to the temporary capture of incident low-energy electron in the effective potential (screened Coulomb and centrifugal potentials) for $l$>0 states. When $D$ approaches to $D_{nl}$, the maxima of both broad and shape resonances within a given $nl$ series increase and their energy positions shift towards lower energies. The half-widths of broad resonances increase when $D$ moves towards $D_{nl}$, while those of shape resonances decrease. The dependence of $g_{ff}(\varepsilon_i, \omega)$ on $\omega$ outside the resonant energy regions is the same as in the Coulomb case, within the resonant region it changes.

**Acknowledgements**

Grants from the National Basic Research Program of China (No. 2017YFA0403200), NSFC (No. 11604197), NSAF (No. U1530142), the Science Challenge Program of China (Nos. TZ2018005, TZ2016005) and the Organization Department of CCCPC are acknowledged.

**Tables and figure captions**

Table 1. Values of the critical screening lengths, $D_{nl}$(a.u.) (Ref.[23]).

| n \ l | 0 | 1 | 2 | 3 | 4 | 5 |
|---|---|---|---|---|---|---|
| 1 | 0.839907 | | | | | |
| 2 | 3.222559 | 4.540956 | | | | |
| 3 | 7.171737 | 8.872221 | 10.947492 | | | |
| 4 | 12.686441 | 14.730720 | 17.210209 | 20.067784 | | |
| 5 | 19.770154 | 22.130652 | 24.984803 | 28.257063 | 31.904492 | |
| 6 | 28.427266 | 31.080167 | 34.285790 | 37.949735 | 42.018401 | 46.458584 |

**Figure captions:**

Fig. 1. (Color online) Non-relativistic free-free absorption Gaunt factors $g_{ff}(\varepsilon_i, \omega)$ for the pure Coulomb potential and field-free cases in the ranges $\varepsilon_i=10^{-8}$-$10^8$ Ry and $\omega=10^{-7}$-$10^7$ Ry. Panel (b) shows $g_{ff}(\varepsilon_i, \omega)$ in the log-log scale.

Fig. 2. (Color online) Non-relativistic free-free absorption Gaunt factors $g_{ff}(\varepsilon_i, \omega)$ for screening lengths in vicinity of $D_{1s}$=0.840 a.u., $D_{2s}$=3.223 a.u. and $D_{2p}$=4.541 a.u. in the ranges $\varepsilon_i=10^{-8}$-$10^2$ Ry and $\omega=10^{-7}$-$10^5$ Ry.

Fig. 3. (Color online) $s$- and $p$-partial wave phase shifts (panels (a) and (c)) of the initial electron and the contributions to the Gaunt factors of the $l$=0, 1 and $l$ >1 waves (panels (b) and (d)) for screening lengths near the critical screening length $D_{2s}$=3.223 a.u. (panels (a) and (b)) and $D_{2p}$=4.541 a.u. (panels (c) and (d)).

Fig. 4. (Color online) Energy dependence of Gaunt factors for $\omega$=10 Ry and screening lengths near the critical screening lengths $D_{3s}$=7.172 a.u. (a), $D_{3p}$=8.872 a.u. (b) and $D_{3d}$=10.947 a.u. (c).

Fig. 5. (Color online) Same as in Fig. 4, but for $D$ in the vicinity of critical lengths $D_{6s}$=28.427 a.u. (a), $D_{6p}$=31.080 a.u. (b), $D_{6d}$=34.286 a.u. (c), $D_{6f}$=37.950 a.u. (d), $D_{6g}$=42.018 a.u. (e) and $D_{6h}$=46.459 a.u. (f).

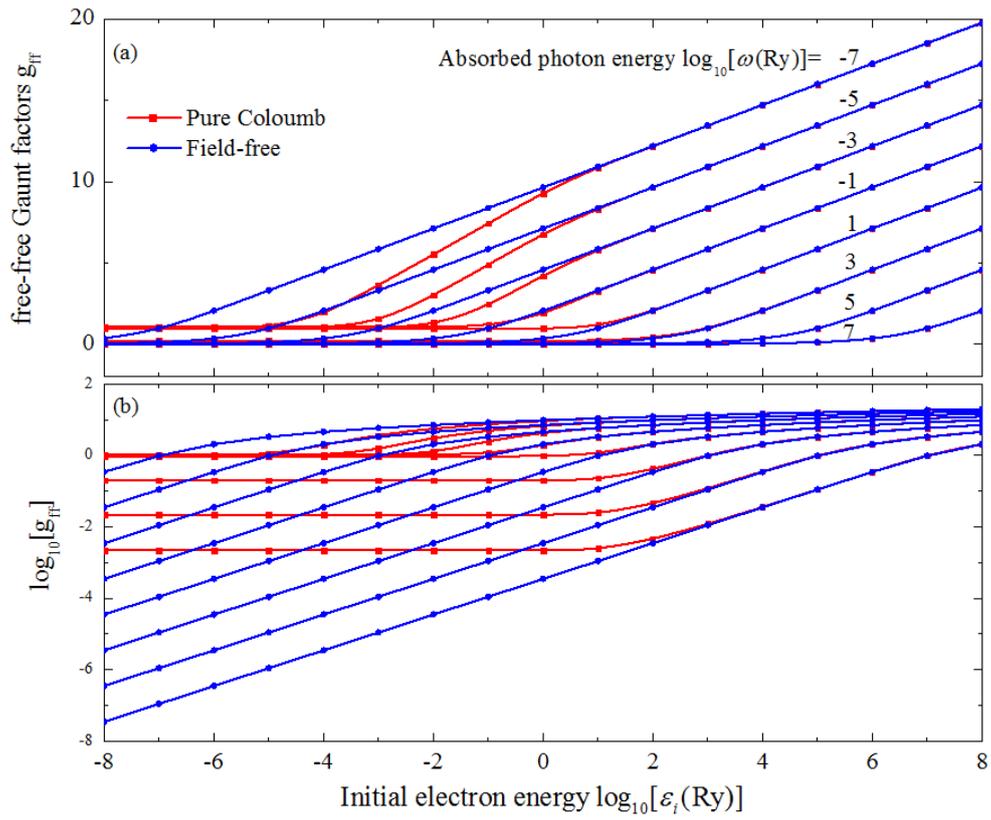

Fig. 1

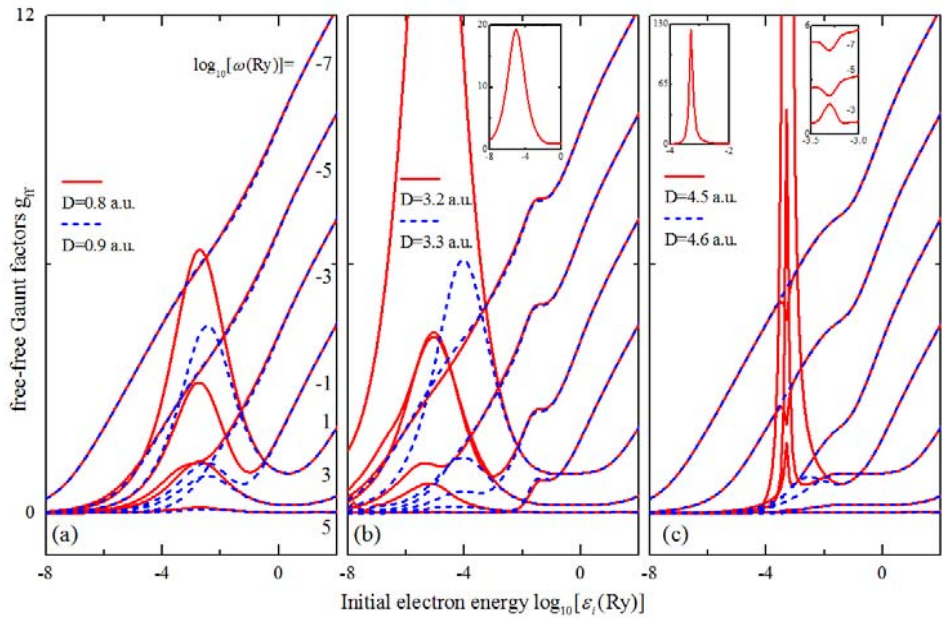

Fig. 2

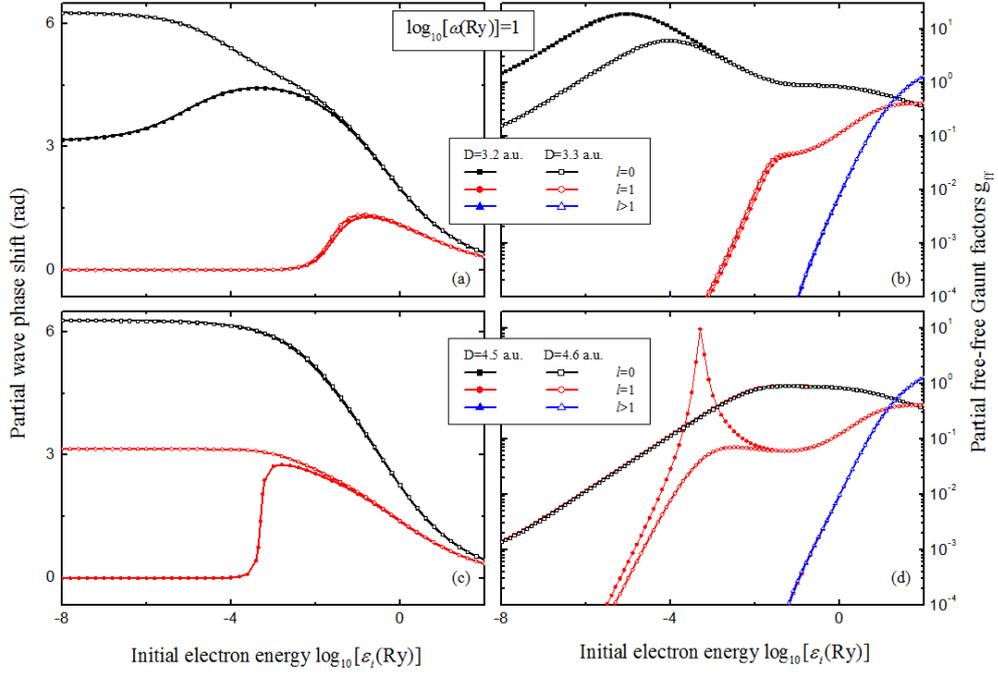

Fig. 3

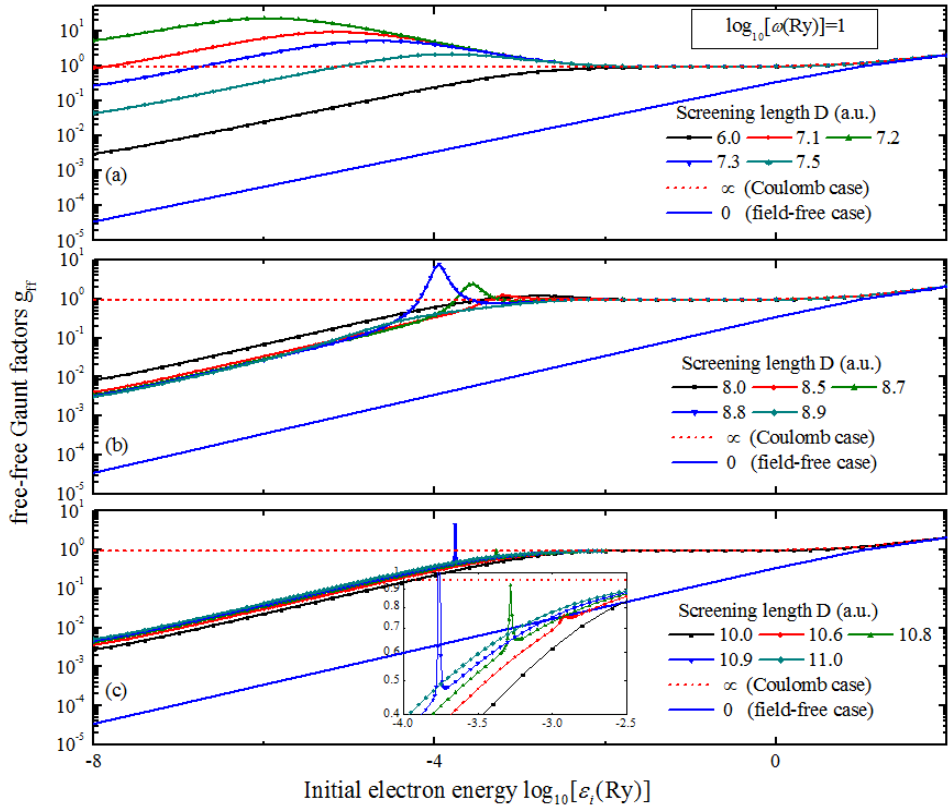

Fig. 4

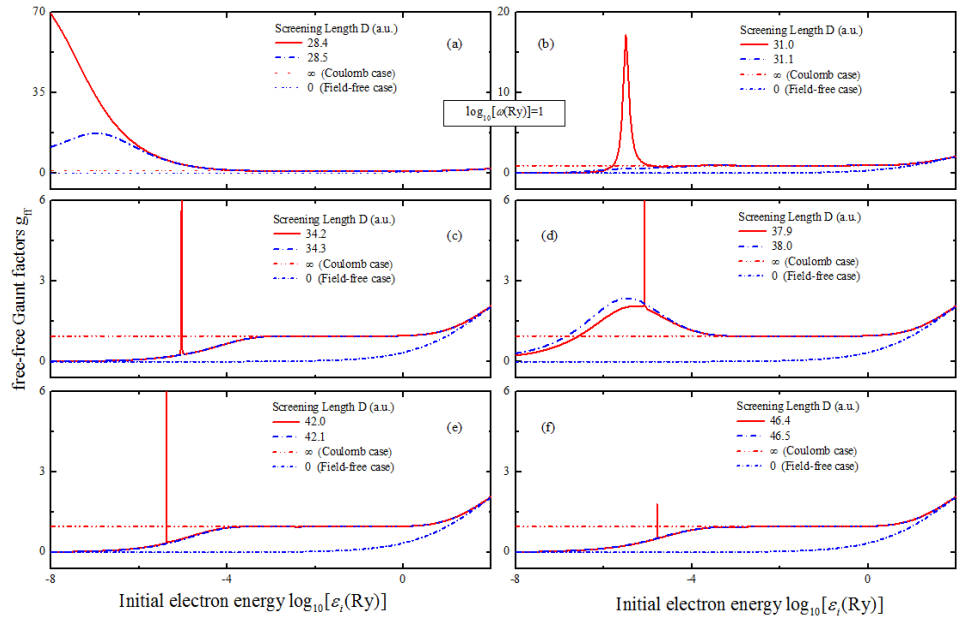

Fig. 5